\documentclass{pasj00}

\begin{document}
\SetRunningHead{A.Tanikawa and T.Fukushige PASJ}{Mass-Loss Timescale of
Star Clusters II.}
\Received{}
\Accepted{}
\title{Mass-Loss Timescale of Star Clusters \\
in an External Tidal Field. \\
II. Effect of Mass Profile of Parent Galaxy}
\author{Ataru \textsc{Tanikawa}$^{1,2}$ and Toshiyuki \textsc{Fukushige}$^{1,3}$}
\affil{$^1$Department of General System Studies, College of Arts and Sciences,\\
University of Tokyo, 3-8-1 Komaba, Meguro-ku, Tokyo 153-8902 \\
$^2$Center for Computational Sciences, University of Tsukuba, \\
1-1-1 Tennodai, Tsukuba, Ibaraki 305-8577 \\
$^3$K\&F Computing Research Co., Chofu, Tokyo 182-0026\\}
\email{tanikawa@ccs.tsukuba.ac.jp}
\maketitle

\begin{abstract}

We investigate the long-term dynamical evolution of star clusters in a
steady tidal field produced by its parent galaxy. In this paper, we
focus on the influence of mass profile of the parent galaxy. The
previous studies were done with the simplification where the parent
galaxy was expressed by point mass. We express different mass profiles
of the parent galaxy by the tidal fields in which the ratios of the
epicyclic frequency $\kappa$ to the angular velocity $\omega$ are
different. We compare the mass-loss timescale of star clusters whose
tidal radii are identical but in parent galaxies with different mass
profile, by means of orbits calculations in fixed cluster potential
and N-body simulations. In this situation, a cluster rotates around
the parent galaxy more rapidly as the parent galaxy has shallower mass
profile. We found that the mass-loss timescale increase $20\%$ and
$50\%$ for the cases that the mass density profile of the parent
galaxies are proportional to $R^{-2}$ and $R^{-1.5}$ where $R$ is the
distance from the galaxy center, compared to the point-mass case, in
moderately strong tidal field. Counterintuitively, a cluster which
rotates around the parent galaxy more rapidly has a longer lifetime.
The increase of lifetime is due to the fact that the fraction occupied
by regular-like orbit increases in shallower profile. Finally, we
derive an evaluation formula for the mass-loss timescale of
clusters. Our formula can explain a property of the population of the
observed galactic globular clusters that their half-mass radii become
smaller as their distances from the galactic center become smaller.

\end{abstract}

\onecolumn

\section{Introduction}

In a star cluster, stars exchange their energy with the other stars
through their encounters, and the whole system approaches to
thermodynamically relaxed state. This process is called two-body
relaxation. In this process, some stars have high energy enough to
escape from the cluster and the system loses its mass. Therefore, the
mass-loss timescale of a cluster is thought to be proportional to its
two-body relaxation time (\cite{Ambart}; \cite{Spit1940}).

However, this is not true for a cluster in an external tidal field,
which was first shown by means of $N$-body simulations in the
Collaborative Experiment (\cite{Heggie1998}). \citet{Fuku2000}
(hereafter, FH) showed that stars which have enough energy to escape
from the cluster (hereafter, ``potential escapers'') can remain inside
the cluster before finding exits on timescale comparable to the
two-body relaxation time of the cluster, and that the timescale
(hereafter, ``escape time delay'') complicates the scaling of
mass-loss timescale with two-body relaxation time. \citet{Baum2001}
and \citet{Baum2003} performed $N$-body simulations in the external
tidal field, and found the mass-loss timescale are proportional to the
half-mass relaxation time to the power of 3/4. \citet{Tani2005}
(hereafter, paper I) also performed $N$-body simulations, and found
that the power depends on the strength of the tidal field and the
power 3/4 shown by \citet{Baum2001} can be observed in the case of
moderately strong tide.  These findings mean that for treatment of
mass loss of clusters in the external tidal field, it is desirable for
orbits of stars to be fully resolved, and $N$-body simulation is more
appropriate than orbit-averaged method, such as Fokker-Planck method.

This is a succeeding paper of Paper I. In this paper, we investigate
the effect of the mass profile of the parent galaxy which produces the
tidal field. In most of previous simulations of star clusters with the
external tidal field (\cite{Giersz94}; \cite{McMillan94};
\cite{Fukushige95}; \cite{Baum2001}; Paper I; \cite{Trenti07}), the
parent galaxy are simplified as point mass.

If we consider the mass profile of the parent galaxy, the equations of
motion of star clusters moving in the circular orbit around the galaxy
are expressed, with tidal approximation, as
\begin{eqnarray}
\displaystyle \frac{d^2{\bf r}_{i}}{dt^2} = - 2 \left( \begin{array}{c}
 0 \\ 0 \\ \omega \\ \end{array} \right) \times \frac{d{\bf r}_{i}}{dt}
- \left( \begin{array}{c} (\kappa^2 - 4 \omega^2) x_{i} \\ 0
 \\ \nu^2 z_{i} \\ \end{array} \right)  - \nabla 
\Phi_{{\rm c},i},\label{eq:motion2}
\end{eqnarray}
where $\omega$ is a angular velocity, $\kappa$ is a epicyclic
frequency, $\nu$ is a vertical frequency, The first term on the
right-hand side in equation (\ref{eq:motion2}) is the Coriolis force,
the second term is a combination of centrifugal and tidal forces, and
the third term is gravitational acceleration from other particles in
the cluster. For the point-mass galaxy, $\kappa=\omega$. Also for the
galaxy with flat rotation ($\rho \propto r^{-2}$), $\kappa = 2\omega$.
In Table \ref{tab:rho}, the approximated relations between mass profile and
$\kappa/\omega$ are summarized. The exact description will be given in
section 2.

\begin{table}
\begin{center}
\caption{Relation among mass density profiles of galaxies, the angular
  velocities of the clusters, and the epicyclic frequency.}
\begin{tabular}{lll}
\hline
\hline
$\rho\propto R^{-\infty}$(point mass) & $\omega \propto R^{-3/2}$ & $\kappa^{2} = \omega^{2}$ \\
$\rho\propto R^{-2}$                  & $\omega \propto R^{-1}$   & $\kappa^{2} = 2\omega^{2}$ \\
$\rho\propto R^{-1.5}$                & $\omega \propto R^{-3/4}$ & $\kappa^{2} = (5/2)\omega^{2}$ \\
$\rho\propto R^{-1}$                  & $\omega \propto R^{-1/2}$ & $\kappa^{2} = 3\omega^{2}$ \\
$\rho\propto R^{0}$                   & $\omega \propto R^{0}$    & $\kappa^{2} = 4\omega^{2}$ \\
\hline
\hline
\end{tabular}
\label{tab:rho}
\end{center}
\end{table}

In this paper, we compare dynamical evolution of the star cluster in
the tidal fields of patent galaxies with four different
$\kappa^2/\omega^2 (=1, 2, 2.5, 3)$, i.e. different mass profile.
Throughout the paper, we set the initial tidal radii identically among
clusters with different $\kappa/\omega$. The tidal radius and that at
the initial time are, respectively, expressed as
\begin{equation}
  r_{\rm t} = \left( \frac{GM}{4\omega^2 - \kappa^2} \right)^{1/3},
  \label{eq:rt}
\end{equation}
and
\begin{equation}
  r_{\rm t,i}=(GM_{\rm i}/(4\omega^2-\kappa^2))^{1/3}, 
  \label{eq:rti}
\end{equation}
where $G$ is the gravitational constant, $M$ and $M_{\rm i}$ are the
cluster mass and that at the initial time. In the condition, the
Coriolis force is relatively stronger for larger $\kappa^2/\omega^2$,
i.e. shallower mass profile. Although several studies set parent
galaxy to have mass density profile other than point mass
(\cite{por2001a}; \cite{por2002}; \cite{por2003}; \cite{Moore1996};
\cite{Baum2003}; \cite{por2001b}; \cite{Yim2002}; \cite{Dehnen2004};
\cite{Capuzzo2005}; \cite{Combes1999}), they aim to study disruption
of star clusters taking place in relatively short timescale.  On
contrary to them, we investigate systematically the effects of the
difference of mass density profile of parent galaxy on long-term
mass-loss timescale of clusters .

The plan of this paper is as follows.  In section 2, we give an exact
description on the equation of motion in the tidal field.  In section
3, we first perform orbital calculation in the fixed potential of star
clusters and we compare the escape time delays of the potential
escapers, which complicates the scaling with two-body relaxation
timescale.  In section 4, we perform $N$-body simulation of clusters
in the tidal field in order to study the influence of mass density
profiles of the parent galaxy on the mass-loss timescale of the star
clusters. In section 5, we apply our simulation results for galactic
globular clusters. Finally we summarize our paper in section 6.

\section{Equation of Motion}

In this section, we derive the equation of motion of stars in a
cluster which moves circularly around a spherically symmetric parent
galaxy with various mass density profiles. We set coordinate system
which moves with the cluster. The origin of the coordinate system is
set at the cluster center, the $x$ axis is oriented in the opposite
direction of the galactic center, and the $z$ axis is along with the
angular velocity vector of the cluster. Then, irrespective of the mass
density profile of parent galaxy, we express the equation of motion of
the cluster stars as
\begin{equation}
  \frac{d^2{\bf r}_i}{dt^2} = - 2 {\bf \Omega} \times \frac{d{\bf
      r}_{i}}{dt} - {\bf \Omega} \times \left\{ {\bf \Omega} \times
  \left[ {\bf r}_i + \left( \begin{array}{c} R_{\rm g} \\ 0 \\ 0
      \\ \end{array} \right) \right] \right\} - \nabla \Phi_{{\rm G},i} -
  \nabla \Phi_{{\rm c},i},
  \label{eq:basiceq}
\end{equation}
where ${\bf r}_i$ is the position vector of $i$-th star, ${\bf
  \Omega}$ is the angular velocity vector of the cluster, $R_{\rm g}$
is the distance between the galactic center and the cluster center,
and $\Phi_{{\rm G},i}$ and $\Phi_{{\rm c},i}$ are, respectively, the
potential of the parent galaxy and the cluster at the position of the
$i$-th star. The first and second terms in equation (\ref{eq:basiceq})
are, respectively, the Coriolis and centrifugal forces on the $i$-th
star.

We consider the situation where the size of the cluster is
sufficiently smaller than the distance from the galactic center to the
cluster. Then, the galactic potential $\Phi_{\rm G}$ can be
expanded in Taylor series around the cluster center, such as
\begin{equation}
  \Phi_{\rm G} \simeq \left. \Phi_{\rm G} \right|_{\bf 0} +
  \left.\frac{\partial \Phi_{\rm G}}{\partial x}\right|_{\bf 0} x +
  \frac{1}{2} \left.\frac{\partial^2\Phi_{\rm G}}{\partial
    x^2}\right|_{\bf 0} x^2 + \frac{1}{2}
  \left.\frac{\partial^2\Phi_{\rm G}}{\partial y^2}\right|_{\bf 0}
  y^2 + \frac{1}{2} \left.\frac{\partial^2\Phi_{\rm G}}{\partial
    z^2}\right|_{\bf 0} z^2,
  \label{eq:taylor1}
\end{equation}
where ${\bf 0}=(0,0,0)$ is the zero vector, the subscripts of ${\bf
  0}$ indicate the galactic potential and its derivatives at ${\bf
  0}$. Since the galactic potential is spherically symmetric,
$(\partial \Phi_{\rm G}/\partial y)_{\bf 0}=(\partial \Phi_{\rm
  G}/\partial z)_{\bf 0}=0$. Furthermore, $(\partial \Phi_{\rm
  G}/\partial x)_{\bf 0}=R_{\rm g}\omega^2$, and $(\partial^2
\Phi_{\rm G}/\partial y^2)_{\bf 0}=(\partial^2 \Phi_{\rm G}/\partial
z^2)_{\bf 0}=\omega^2$, where $\omega$ is the angular speed of the
cluster around the galactic center, i.e. ${\bf
  \Omega}=(0,0,\omega)$. Finally, equation (\ref{eq:taylor1}) is
rewritten as
\begin{equation}
  \Phi_{\rm G} \simeq \left. \Phi_{\rm G} \right|_{\bf 0} + R_{\rm
    g} \omega^2 x + \frac{1}{2} \left.\frac{\partial^2\Phi_{\rm
      G}}{\partial x^2}\right|_{\bf 0} x^2 + \frac{1}{2} \omega^2
  y^2 + \frac{1}{2} \omega^2 z^2.
  \label{eq:taylor2}
\end{equation}
Substituting equation (\ref{eq:taylor2}) into equation
(\ref{eq:basiceq}), the equation of motion of the $i$-th star is
expressed as
\begin{equation}
  \frac{d^2{\bf r}_i}{dt^2} = - 2 {\bf \Omega} \times \frac{d{\bf
      r}_i}{dt} - \left( \begin{array}{c} (\kappa^2 - 4 \omega^2)x_i
    \\ 0 \\ \omega^2z_i \\ \end{array} \right) - \nabla \Phi_{{\rm
      c},i},
  \label{eq:eom}
\end{equation}
where 
\begin{equation}
\kappa^2= \left. \frac{\partial^2 \Phi_{\rm G}}{\partial x^2}
\right|_{\bf 0} + 3\omega^2,
\label{eq:kappa}
\end{equation}
and $\kappa$ is so-called epicyclic frequency.

The epicyclic frequency, $\kappa$, and angular speed of the cluster,
$\omega$, are determined by the mass density profile of the galaxy,
$\rho_{\rm G}$, and the distance between the galactic center and the
cluster center, $R_{\rm G}$. The angular speed, $\omega$, is expressed
as
\begin{equation}
  \omega^2 = \frac{GM_{\rm g}(0)}{R_{\rm g}^3} = \frac{4}{3}\pi G
  \overline{\rho_{\rm G}}(0),
  \label{eq:omega}
\end{equation}
where $M_{\rm g}(0)$ and $\overline{\rho_{\rm G}}(0)$ are,
respectively, the galactic mass and average mass density within the
galactic radius, $R_{\rm g}$. Since the mass density profile is
spherically symmetric, the galactic potential, $\Phi_{\rm G}$, is
expressed as
\begin{equation}
  \Phi_{\rm G} = - 4\pi G\left[ \frac{1}{x+R_{\rm g}}
    \int^{x}_{-R_{\rm g}} \rho_{\rm G}(x') \left(x'+R_{\rm g}\right)^2
    dx' + \int^{\infty}_x \rho_{\rm G}(x')\left( x'+ R_{\rm g} \right)
    dx' \right].
  \label{eq:phig}
\end{equation}
Differentiating partially equation (\ref{eq:phig}) with respect to
$x$, we obtain
\begin{equation}
  \frac{\partial^2 \Phi_{\rm G}}{\partial x^2} = - \frac{2GM_{\rm
      g}(x)}{(x+R_{\rm g})^3}+4\pi G\rho_{\rm G}(x),
\end{equation}
and therefore,
\begin{equation}
  \left. \frac{\partial^2 \Phi_{\rm G}}{\partial x^2} \right|_{\bf 0}
  = -2\omega^2 + 4 \pi G \rho_{\rm G}(0).
  \label{eq:d2phi}
\end{equation}
Substituting equation (\ref{eq:d2phi}) into equation (\ref{eq:kappa}),
we obtain the ratio of the epicyclic frequency to the angular speed,
such as
\begin{equation}
  \frac{\kappa^2}{\omega^2} = 1 + \frac{4 \pi G \rho_{\rm
      G}(0)}{\omega^2} = 1 + 3 \frac{\rho_{\rm
      G}(0)}{\overline{\rho_{\rm G}}(0)},
  \label{eq:kwratio}
\end{equation}
where the second equality in equation (\ref{eq:kwratio}) comes from
equation (\ref{eq:omega}).

From equation (\ref{eq:kwratio}), we can confirm the relation between
the mass density profile of the galaxy and the ratio of $\kappa^2$ to
$\omega^2$ in table \ref{tab:rho}. Since $\rho_{\rm G}(x)=0$ in the
case of $\rho_{\rm G}(x) \propto R^{-\infty}$, we can see immediately
$\kappa^2=\omega^2$. We consider the case:
\begin{equation}
    \rho_{\rm G}(x) = C (x+R_g)^{-k} \; (0 \le k \le 2).
    \label{eq:rhog}
\end{equation}
Therefore, the average mass density within the galactic radius,
$R_{\rm g}$, is expressed as
\begin{equation}
  \overline{\rho_{\rm G}}(0) = \frac{\int^0_{-R_{\rm g}}4\pi (x+R_{\rm
      g})^2\rho_{\rm G}dx}{4\pi R_{\rm g}^3 / 3} =
  \frac{3}{3-k}C(x+R_{\rm g})^{-k},
  \label{eq:rhobar}
\end{equation}
where the second equality is obtained by using equation
(\ref{eq:rhog}). Substituting equation (\ref{eq:rhog}) and
(\ref{eq:rhobar} into equation (\ref{eq:kwratio}), equation
(\ref{eq:kwratio}) is rewritten as
\begin{equation}
  \frac{\kappa^2}{\omega^2} = 4 - k \; (0 \le k \le 2).
\end{equation}

The tidal radius is defined as the distance from the cluster center to
a Lagrange point, and can be expressed by equation (\ref{eq:rt}). In
this paper, we investigate the cluster evolution by parameterizing the
mass density profile of the galaxy, $\rho_{\rm G}$, and the initial
tidal radius given by equation (\ref{eq:rti}), not the mass density
profile of the galaxy, $\rho_{\rm G}$, and the distance between the
galactic center and the cluster center, $R_{\rm g}$. We regard the
initial tidal radius, $r_{\rm t,i}$ as one of the most important
parameter for the mass-loss timescale of the cluster, since in paper I
we found that the longer the mass-loss timescale of the cluster
becomes, the larger the initial tidal radius is even if the mass
density profile of the galaxy is equal.

We consider the situation where the initial tidal radius, $r_{\rm
  t,i}$, i.e. $(4\omega^2-\kappa^2)$, is kept constant. From equation
(\ref{eq:kwratio}), the angular speed, $\omega$, becomes larger, as
the mass density profile of the galaxy becomes shallower, or
$\rho_{\rm G}(0)/\overline{\rho_{\rm G}}(0)$ becomes smaller. In the
equation of motion (\ref{eq:eom}), the Coriolis force becomes larger
as the mass density profile of the galaxy becomes shallower.

As an example, we investigate the difference of the Coriolis forces
between point-mass galaxy ($\rho_{\rm G}(x)=0,x>-R_{\rm G}$) and
isothermal sphere galaxy ($\rho_{\rm G}(x)=3M_{\rm G}(0)/(4\pi R_{\rm
  G} (x+R_{\rm G})^2),x>-R_{\rm G}$). The values of
$\kappa^2/\omega^2$ in point-mass galaxy and isothermal sphere galaxy
are $1$ and $2$, respectively. Then, Coriolis force increases by $20$
\% from the case of point-mass galaxy to that of isothermal sphere
galaxy. The Coriolis force changes angular momentum of cluster stars
around cluster center. The difference of the Coriolis force may affect
escape of cluster stars.

\section{Orbital Calculation of Potential Escaper in Fixed Potential}

In this section, we perform orbit calculations in the fixed and smooth
cluster potential in order to investigate the escape time delay of the
potential escapers. Due to their mutual encounters, stars gain enough
energy to escape from the clusters and become the potential escaper.
Such process takes place in two-body relaxation timescale.  In the
tidal field, the potential escapers can escape only through small
apertures around the two Lagrangian points, and can remain inside the
clusters before the escape on timescale comparable to two-body
relaxation time (FH). Therefore, such escape time delay
can influence on the escape timescale or the mass-loss timescale of
the star clusters.

\subsection{Method}

We calculate the orbits of the potential escapers in a fixed and
smoothed potential due to cluster stars plus the steady tidal field of
the parent galaxy. The equation of motion of the potential escapers is
described in equation (\ref{eq:eom}).  For the cluster potential,
$\Phi_{{\rm c},i}$, we use $W_{0}=3$ King model, where $W_{0}$
indicates initial dimensionless central potential. In order to compute
the acceleration from the King model, $\nabla \Phi_{{\rm c},i}$, in
equation (\ref{eq:eom}), we use cubic spline interpolation between
grid data obtained by a numerical solution (e.g. \cite{Press}).

Regardless of the mass density profile of the parent galaxy, we set
the initial tidal radius of the cluster, $r_{\rm t,i}$ to be equal to
the radius beyond which the density is zero in the $W_{0}=3$ King
model, denoted by $r_{\rm kg}$. The value of $r_{\rm kg}$ is $3.13$,
where we use standard units (\cite{Heggie1986}), such that $M_{\rm i}
= G = -4 E_{\rm c} = 1$, where $E_{c}$ is the initial total energy
within the cluster. When $r_{\rm t,i}=r_{\rm kg}=3.13$, the values of
$\kappa$ and $\omega$ are determined for each given ratio
$\kappa^2/\omega^2$, using equation (\ref{eq:kwratio}) and
(\ref{eq:rt}). We performed the orbital calculations for 4 models in
which the parent galaxy has mass density profiles of
$\kappa^2/\omega^2=1,2,2.5,3$. The values of $\kappa^2$ and $\omega^2$
are showed in table \ref{tab:init}. The second right column shows
representative mass density profiles of the parent galaxy in each
ratio of $\kappa^2/\omega^2$, such as single power law models.

We determine the escape time delay as the duration between $t=0$ and
the time when escape condition, $|x_{i}|>r_{\rm t}$, is satisfied.
Note that the tidal radius, $r_{\rm t}$, remains the initial tidal
radius, $r_{\rm t,i}$, during the calculation, since clusters do not
lose the mass in this calculations.

We calculate sets of orbits of the potential escapers for several values of
relative energy excess,
${\hat{E}}_{{\rm pot},i} = (E_{{\rm pot},i} - E_{\rm crit}) / |E_{\rm crit}|$,
where $E_{{\rm pot},i}$ is the energy of $i$-th potential escaper, given
by
\begin{equation}
E_{{\rm pot},i}(t) = \frac{1}{2} ({v_{{\rm x}i}}^2 + {v_{{\rm y}i}}^2 +
 {v_{{\rm z}i}}^2) + \Phi_{{\rm c},i} -\frac{1}{2} (\kappa^2 - 4
 \omega^2) {x_{i}}^2 + \frac{1}{2}\omega^2 {z_{i}}^2,
\end{equation}
where $v_{{\rm x}i}$, $v_{{\rm y}i}$, and $v_{{\rm z}i}$ are $x$, $y$,
and $z$ components of the velocity of $i$-th particle, respectively.

We integrate the orbits of the stars by means of adaptive stepsize
control for a forth-order Runge-Kutta method (e.g. \cite{Press}). The
integration error in energy is about $9 \times 10^{-5}$ at maximum
(${\hat{E}}_{{\rm pot},i}=0.03$ at $t = 3 \times 10^7$ in standard
units in the shallowest mass density profile of parent galaxy,
$\kappa^2/\omega^2=3$).

\subsection{Results}

In this subsection, we present the results of the orbital calculations
of the potential escapers. In section 3.2.1, we investigate the
difference of orbit families in planar orbits ($z=0$ plane) among
parent galaxies with different mass density profiles. As the profile
of the parent galaxy is shallower, more potential escapers move in
regular-like orbit, and have larger escape time delay. In section
3.2.2, we show that this is also applicable to non-planar orbit
families. In section 3.2.3, we investigate the distribution of the
escape time delay in order to see whether the difference of orbit
families affects the escape rate of potential escapers.

\subsubsection{Orbital families in planar orbits}

At first, our investigation is confined to the orbits of potential
escapers on $z=0$ plane for simplicity of the analyses. If we give the
potential escapers initial conditions where $z_{i}=0$ and
${\dot{z}}_{i}=0$, their orbits are confined to $z=0$ plane forever
shown in equation (\ref{eq:eom}).

\begin{figure}
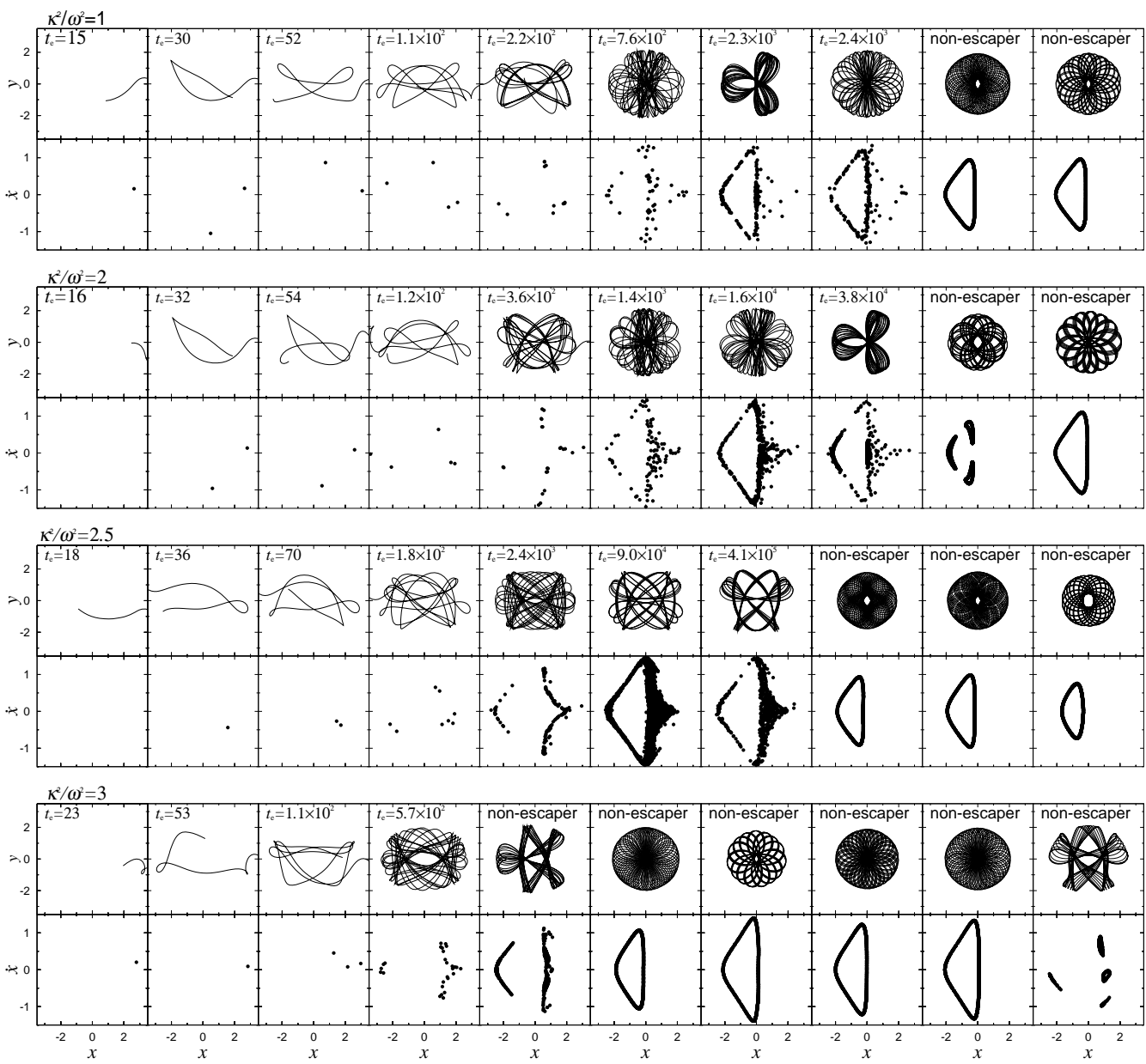

\begin{center}
\FigureFile(160mm,100mm){random_orbit_xxdot.eps}
\end{center}
\caption{Sets of orbits and surfaces of section of potential
  escapers. We present the orbits of the potential escapers only at $t
  < 5 \times 10^2$, and surfaces of section of the potential escapers
  until $t < 10^6$. In the surfaces of section, points represent the
  phase space $(x, \dot{x})$ of the potential escapers when they cross
  $x$ axis in the direction of $\dot{y}>0$. In each model, they are
  arrayed in small order of escape time delay from left to right.}
\label{fig:analy_orbit}
\end{figure}

In figure \ref{fig:analy_orbit}, we illustrate orbits and surfaces of
section for 10 potential escapers with various escape time, $t_e$, for
each model of $\kappa^2/\omega^2=1,2,2,5,3$.  We plot the orbit until
the escape or $t=500$ for non-escaper, and the surfaces of section
until the escape or $t = 10^6$ for non-escaper.  Here and hereafter,
potential escapers which are still in clusters when orbit calculations
are terminated are called ``non-escaper''.  We define the surface of
section as $(x, \dot{x})$ at which the potential escapers cross $x$
axis in the direction of $\dot{y}>0$.

At initial, we distribute $1000$ potential escapers of
${\hat{E}}_{{\rm pot},i}=0.03$ uniformly in $(x, y, \dot{x}, \dot{y})$
phase space, and calculate the escape time delays.  We sort the $1000$
potential escapers in small order of the escape time delay. We chose
$k$-th potential escapers, where $k=50+100(i-1) (i=1,2,\ldots, 10)$
(hereafter, we call $k(i)$-th potential escapers ``$i$-th potential
escapers''), and show the orbit and surface of section of $i$-th
potential escaper in $i$-th panel of each model in figure
\ref{fig:analy_orbit} from left to right. As for the non-escapers, the
selection are done at random.

In the models of two steeper mass density profiles
($\kappa^2/\omega^2=1$ and 2), the orbits become regular-like
gradually from left to right. In both models, the $7-10$th potential
escapers looks like regular. However, in the models of two shallower
mass density profile galaxies ($\kappa^2/\omega^2=2.5$ and 3), the
orbits are seen to become regular-like suddenly at the $5$th potential
escapers.

In all models, there are potential escapers whose surfaces of section
are confined to small phase space, and they correspond to
non-escapers. We call these orbits ``regular orbits''. In the model of
the shallow mass density profile ($\kappa^2/\omega^2=2.5$), some
orbits seem regular, but their surfaces of section occupy larger space
than regular orbits, and eventually escape from the clusters, such as
$6$th and $7$th potential escapers. We call these orbits
``semi-regular orbits''.

\begin{figure}
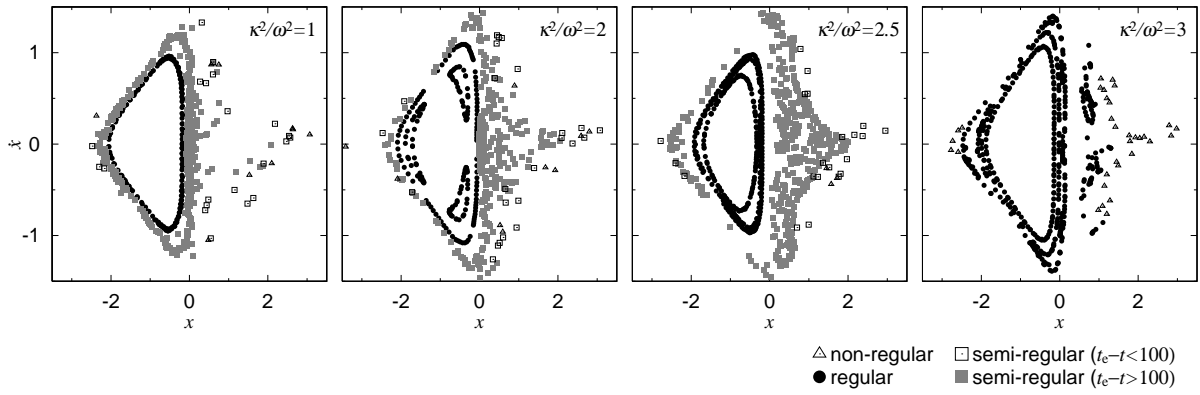

\begin{center}
\FigureFile(160mm,100mm){xxdot_time_divide_sum.eps}
\end{center}
\caption{For each model, surfaces of section in figure
  \ref{fig:analy_orbit} are put together. Surface of section is
  divided into four kinds, regular orbit (black circle), non-regular
  orbit (triangle), and semi-regular orbit (gray and white
  squares). The colors of the squares are divided, whether the
  potential escapers cross $x$ axis in the direction of $\dot{y}>0$
  less (white) or more (gray) than $100$ time units before escape.}
\label{fig:xxdot_time_group}
\end{figure}

We consider the difference among the regular, semi-regular, and the
other orbits. In figure \ref{fig:xxdot_time_group} we put together
surface of section of the $1$st, $2$nd, $\ldots$, $10$th potential
escapers in each model. The colors and shapes in surface of section
are divided according to non-regular (white triangles), semi-regular
(white or gray squares), and regular orbits (black circles), where we
define semi-regular orbit as the orbit of potential escapers which
have escape time delay, $t_{\rm e}>100$ and non-regular orbit as
escape time delay, $t_{\rm e}<100$. The surfaces of section of
semi-regular orbits, whose escape time delay is $t_{\rm e}>100$, are
divided into two colors, white and gray. The difference of the colors
shows not the difference among the potential escapers, but that of the
time when the potential escapers cross $x$ axis in the direction of
$\dot{y}>0$. If the potential escapers cross $x$ axis in the direction
of $\dot{y}>0$ less and more than $100$ time units before they escape
from the cluster, the colors of the surfaces of section are,
respectively, white and gray.

The region in the surface of section occupied by the non-regular and
semi-regular orbits is different from that of regular orbits in each
model. The region in the surface of section for non-regular orbits is
similar to that of semi-regular orbits just before escape, i.e. that
of gray squares, and different from that of semi-regular orbits not
just before escape, i.e. that of white squares. From the above, we
guess that the escape time delay of any potential escaper depends on
their initial phase space.

\begin{figure}
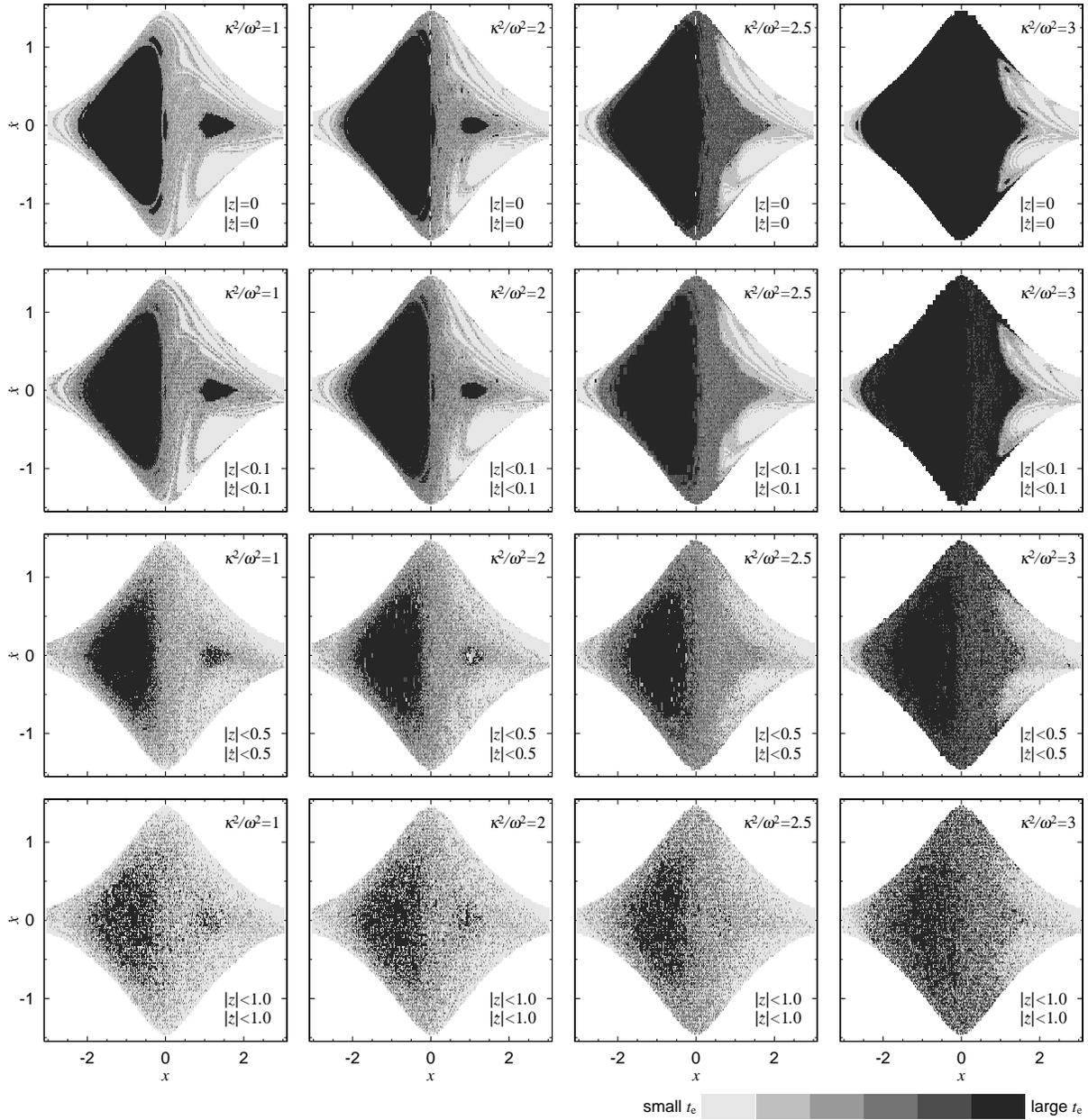

\begin{center}
\FigureFile(160mm,100mm){rev_gray.eps}
\end{center}
\caption{Escape time delay of potential escapers as a function of
  their initial phase space $(x,\dot{x})$. The color of initial phase
  space in each panel becomes deep from gray to black in order of
  escape time delay, $t_{\rm e}<10^2$, $10^2<t_{\rm e}<10^3$,
  $10^3<t_{\rm e}<10^4$, $10^4<t_{\rm e}<10^6$, and non-escaper. The
  conditions about $|z|$ and $|\dot{z}|$ show initial phase space of
  potential escapers, and their $(z,\dot{z})$ is distributed in the
  range at random.}
\label{fig:mozaic}
\end{figure}

Figure \ref{fig:mozaic} shows the escape time delay of potential
escapers against the initial phase space $(x,\dot{x})$.  The uppermost
4 panels show the potential escapers whose initial phase is $y=0$,
$z=0$, $\dot{y}>0$, and $\dot{z}=0$. The points are colored according
to the escape time delay, $t_{\rm e}$, of the potential
escapers. Black regions show non-escapers, and the lightest regions
show the potential escapers of $t_{\rm e}<10^2$. The escape time delay
depends on the initial phase of the potential escapers.  As the mass
density profile of the parent galaxy is shallower from
$\kappa^2/\omega^2=1,2$ to $\kappa^2/\omega^2=2.5$, the phase volume
for $10^2 < t_{\rm e} < 10^4$ becomes smaller, and that of $t_{\rm
  e}>10^4$ becomes larger. Furthermore, as the mass density profile of
the parent galaxy becomes shallower from $\kappa^2/\omega^2=2.5$ to
$3$, the phase space for $10^4 < t_{\rm e} < 10^6$ is replaced by that
of non-escapers. The phase space of $t_{\rm e} < 10^2$ is similar in
all models.

In summary, we find that there is a larger fraction of regular-like
orbital families in the parent galaxy with shallower mass density
profiles.  The difference in the orbital families among galaxy models
brings the structure in initial phase space about their escape time
delay shown in uppermost panel of figure \ref{fig:mozaic}.

\subsubsection{Orbital families in non-planar orbits}

Figure \ref{fig:mozaic} shows the escape time delay against the
initial phase $(x, \dot{x})$ also for non-planar orbits. The points
are colored according to the escape time delay of potential escapers.
In each panels the escape time delay of the potential escapers whose
initial positions and velocities of $z$-component are in limited
ranges are only plotted, and whose initial positions and velocities of
$y$-component are $y=0$, and $\dot{y}>0$. The $z$ components increase
from top panels to bottom panels. The structure in initial phase
against the escape time delay seen in z=0 planer orbit can be also
seen in other non-planer orbit, though the structure becomes less fine
as the $z$ components increases.

\subsubsection{Distribution of escape time delay}

\begin{figure}
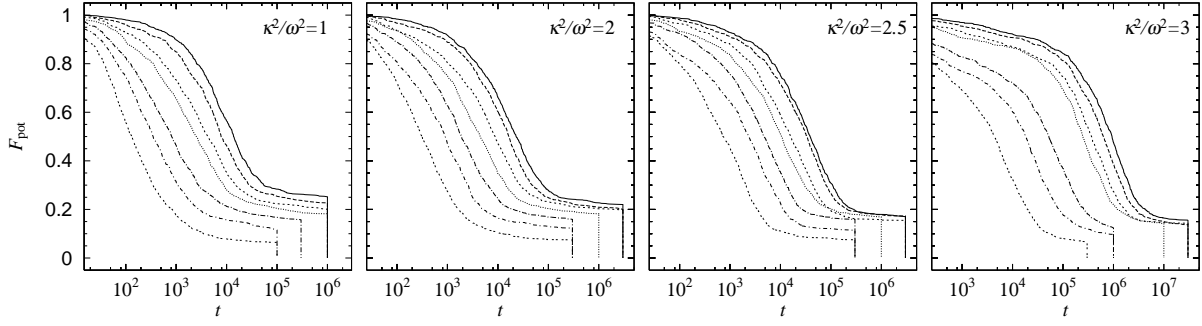

\begin{center}
\FigureFile(160mm,100mm){te_dstr_kg_all.eps}
\end{center}
\caption{Distribution of escape time delay of potential escapers. Mass
  density profile of parent galaxy becomes shallower from left panel
  to right one. In each panel, the potential escapers have, from left
  to right, relative energy excess ${\hat{E}}_{{\rm
      pot},i}=0.24,0.16,0.12,0.08,0.06,0.04$ and $0.03$.}
\label{fig:te_dstr_kg_all}
\end{figure}

Figure \ref{fig:te_dstr_kg_all} shows fractions, $F_{\rm pot}$, of
potential escapers which do not escape at a given time $t$.  The
curves are drawn, from left to right, in order of large initial
relative energy excess, ${\hat{E}}_{{\rm pot},i} =
0.24,0.16,0.12,0.08,0.06,0.04$ and $0.03$.  We spatially distribute
them in the same way as primordial population in $W_{0}=3$ King model
immersed in a tidal field of parent galaxy.  The vertical lines at the
right side show the times when the integrations are terminated. A
common feature in all models is that the potential escapers of larger
relative energy excess, ${\hat{E}}_{{\rm pot},i}$, have smaller escape
time delay and smaller fraction of non-escapers. The feature in the
case of the point mass ($\kappa^2/\omega^2=1$) agrees with the result
of FH.

\begin{figure}
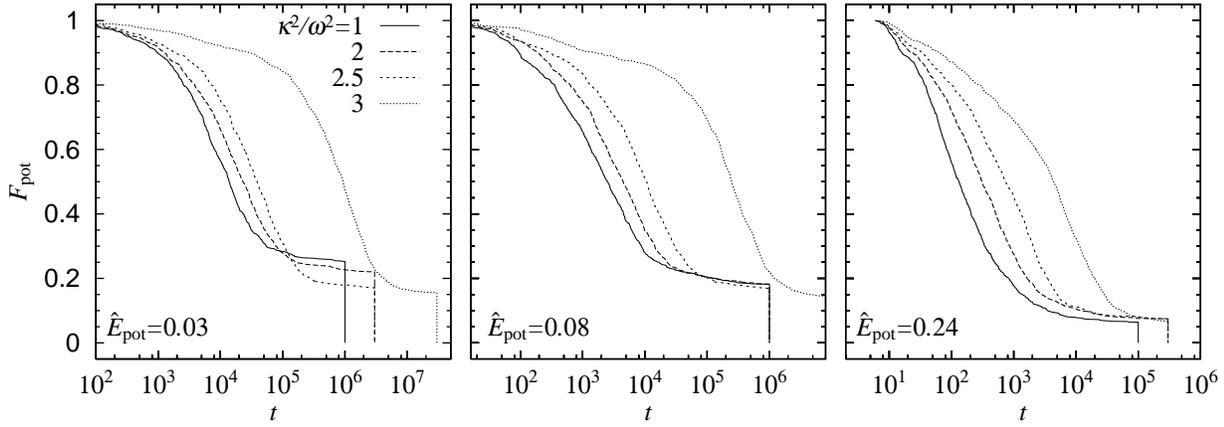

\begin{center}
\FigureFile(160mm,100mm){te_dstr_kg.eps}
\end{center}
\caption{Distribution of escape time delay of potential
 escapers. Relative energy excesses of potential escapers are
 $0.03,0.08$ and $0.24$, from left panel to right one.}
\label{fig:te_dstr_kg_0.03}
\end{figure}

Figure \ref{fig:te_dstr_kg_0.03} shows the fractions, $F_{\rm pot}$,
of potential escapers which do not escape at a given time $t$.  We
plot together the fractions, $F_{\rm pot}$, of all models of
$\kappa^2/\omega^2$ in each relative energy excess, ${\hat{E}}_{{\rm
pot},i}=0.03, 0.08$ and $0.24$. The potential escapers in parent
galaxy with shallower mass density profile escape more slowly, which
agrees with the difference of orbit families described in section
3.2.1.

We try to explain the difference of escape time delay among different
${\hat{E}}_{{\rm pot},i}$ and $\kappa^2/\omega^2$ based on phase space
flux of orbits of potential escapers. We define phase space flux near
Lagrange points as amount of phase space which flows near Lagrange
points in the outward direction with respect to the cluster per time
unit. If potential escapers are equally present in every phase space,
the number of escapers from the cluster should be proportional to the
above phase space flux. In FH, which is the same as model of
$\kappa^2/\omega^2=1$, the phase space flux may explain the dependence
of escape time delay on relative energy excess.

The phase space flux, ${\cal F}$, per unit energy is expressed (\cite{MacKay}) by
\begin{equation}
\displaystyle {\cal F} = 2 \int_{\dot{x'}>0} \delta \left[ \phi' +
						     \frac{1}{2}
						     ({\dot{x'}}^2 +
						     {\dot{y'}}^2 +
						     {\dot{z'}}^2) - E
						   \right] \dot{x'}
						     d\dot{x'} d\dot{y'}
						     d\dot{z'} dy' dz',
						     \label{eq:flux}  
\end{equation}
where $x', y', z', \dot{x'}, \dot{y'}$, and $\dot{z'}$ are coordinate
and velocity when the origin is located in a Lagrange point which have
$x=r_{\rm t}$, and $\phi'$ is obtained by expanding the effective
potential, $\phi$, around the Lagrange point to second order. $\phi$ and
$\phi'$ is respectively expressed by  
\begin{equation}
\displaystyle \phi = - \frac{1}{2} (\kappa^2 - 4 \omega^2) x^2 +
 \frac{1}{2} \omega^2 z^2 + \Phi_{\rm c}, \label{phi}
\end{equation}
where $\Phi_{\rm c}=-GM/r_{\rm t}$, and
\begin{equation}
\phi' - E_{\rm crit} = \frac{3}{2} (\kappa^2 - 4 \omega^2) {x'}^2 -
 \frac{1}{2} (\kappa^2 - 4 \omega^2) {y'}^2 - \frac{1}{2} (\kappa^2 - 5
 \omega^2) {z'}^2. \label{eq:phieff}
\end{equation}
Since there are two Lagrange points, we must add coefficient of $2$ to
integral. In equation (\ref{eq:flux}), $\phi'$ is evaluated at
$x'=0$. We integrate equation (\ref{eq:flux}), and obtain
\begin{equation}
\displaystyle {\cal F} = \frac{4 \pi^2 (E - E_{\rm crit})^2}{\sqrt{(4
 \omega^2 - \kappa^2)(5 \omega^2 - \kappa^2)}}. 
\end{equation}
A familiar calculation shows that the phase space volume per unit energy
is expressed by 
\begin{equation}
{\cal V} = 4 \pi \int \sqrt{2(E-\phi)}d^3{\bf r},
\end{equation}
where the integration is done over the inside cluster. This does not
depend sensitively on $E$ in the vicinity of $E=E_{\rm crit}$, and so we
evaluate it by a Monte Carlo technique. The phase space fluxes and
volumes per energy in ${\hat{E}}_{\rm pot}=0.03$ and the timescale of
phase space flux, $t_{\rm ph} = {\cal V}/{\cal F}$, are summarized in
table \ref{tab:phsp}.

Figure \ref{fig:te_tescale_kg_all} shows the fraction, $F_{\rm esc}$,
of escapers which do not escape at a given time scaled by the
timescale of the phase space flux, $t/t_{\rm ph}$. We exclude
non-escapers from figure \ref{fig:te_dstr_kg_all}. As seen in figure
\ref{fig:te_tescale_kg_all}, the $F_{\rm esc}$ curves are in good
agreement in three steeper mass density profiles
($\kappa^2/\omega^2=1,2,2.5$). The result in the case of point-mass
galaxy agrees with that of
FH. We can see that escape time delay, $t_{\rm e}$, is proportional to
the timescale of the phase space flux, $t_{\rm ph} \propto (E - E_{\rm
  crit})^{-2} \propto {\hat{E}}^{-2}$ in the case of steeper mass
density profile of parent galaxy ($\kappa^2/\omega^2 \le
2.5$). However, in the shallowest mass density profile
($\kappa^2/\omega^2=3$), the agreement becomes worse.

In Figure \ref{fig:te_tescale_kg_0.03}, the fraction, $F_{\rm esc}$ of
${\hat{E}}_{{\rm pot},i}=0.03$ for all models are plotted together
against time scaled by the timescale of the phase space flux,
$t/t_{\rm ph}$.  Agreement in the curves is not good, and escape time
delay scaled by the timescale of the phase space flux, $t_{\rm
  e}/t_{\rm ph}$, are longer as the mass density profile of the parent
galaxy becomes shallower. Figure \ref{fig:te_tescale_kg_0.03} shows
that the timescale of the phase space flux may not well explain the
dependence of the escape time delay the on mass density profile of the
parent galaxy.

\section{$N$-body Simulations}

In this section, we present results of $N$-body simulations of star
clusters, and investigate how the dependence of escape time delay on
mass density profile of parent galaxy influence the mass-loss
timescale of clusters.  In section 4.1, we describe the simulation
method. In section 4.2, we show the results, and find that mass-loss
timescale becomes larger as mass density profile of parent galaxy is
shallower.

\subsection{Simulation method}

We investigate mass loss of star clusters in an external tidal field
by means of $N$-body simulations. The equation of motion is described
in equation (\ref{eq:eom}). In our simulations we use a softened
gravitational potential, and the third term of equation
(\ref{eq:eom}) is expressed as 
\begin{equation}
\displaystyle - \nabla \Phi_{{\rm c},i} = - \sum^{N}_{j=1,j \neq i}
 \frac{G m_{j} ({\boldsymbol{r}}_{i} -
 {\boldsymbol{r}}_{j})}{({|{\boldsymbol{r}}_{i} -
 {\boldsymbol{r}}_{j}|}^2 + \varepsilon^2)^{3/2}}, \label{eq:phici}
\end{equation}
where $m_{j}$ is mass of $j$-th particle and $\varepsilon$ is a
softening parameter. We set the softening parameter, $\varepsilon$, as
$1/32$, where we use the standard units, $M_{\rm i} = G = -4 E_{\rm c}
= 1$, as section 2.

We use $W_{0}=3$ King's models \citep{King1966} to generate initial
distribution of star clusters. We perform eight sets of simulations of
star clusters with different initial tidal radii, $r_{\rm t,i}$, and
different mass density profiles of parent galaxies,
$\kappa^2/\omega^2$, as summarized in table \ref{tab:init}.  The
number of particles used for runs are $N=2^{i}$ in the range described
in table \ref{tab:init}. All particles have the same mass, $m=M_{\rm
  i}/N$. We perform five runs whose realization of particle
distribution are different for each $N$ when $N \le 8192$, and one run
when $N \ge 16384$, except for model $r_{\rm t,i}/r_{\rm kg}=2.2$ and
$\kappa^2/\omega^2=3$. For this model, five runs when $N \le 4096$,
and one run when $N \ge 8192$.

The simulation code is the same as that used in Paper I.  We perform
numerical integrations of equation (\ref{eq:motion2}) using a
leap-frog integration scheme with shared and constant timestep. The
stepsize, $\Delta t$, is set to be as $1/64$ in models $r_{\rm
t,i}/r_{\rm kg}=1.0$ and $\kappa^2/\omega^2=1,2,2.5$, and as $1/128$
in the other models. We used the Barnes-Hut tree algorithm
\citep{Barnes1986} on GRAPE-5 \citep{Kawai2000}, a special-purpose
computer designed to accelerate $N$-body simulations.  We use only the
dipole expansion and the opening parameter $\theta = 0.5$. It spend
about 500 CPU hours completing the longest run, $N=32768$ in models
which have larger tidal radii ($r_{\rm t,i}/r_{\rm kg}=2.2$) and the
steepest mass density profile of parent galaxy
($\kappa^2/\omega^2=1$). For smaller $N$ simulations, the force
calculation is done by direct summations (when $N \leqq 4096$) and on
host computer (without GRAPE-5, when $N \leqq 512$).

Contrary to the standard star cluster simulations, our simulation uses
a softened gravitational potential and a leap-flog integrator with a
relatively large stepsize. Also, the force calculation is performed
with the tree algorithm. We adopt these approaches, since we partly
use mass-loss timescale obtained in paper I, which modeled mass
density profile as the point-mass one ($\kappa^2/\omega^2=1$). As
discussed in paper I, the approaches that we adopt do not influence
the results concerning escape from the cluster.

\subsection{Results}

Figure \ref{fig:mass_evo} shows evolutions of the total mass for all
models. The curves indicate the decrease in mass of the cluster
defined by a tidal boundary. We define geometrically the cluster
member as all stars within the tidal radius from the center of mass of
the cluster, which is expressed as equation (\ref{eq:rt}). Since $M$
depends on $r_{\rm t}$, itself, some iteration is usually required. We
remove stars when they escape far enough (more than $4096$ length
units from the coordinate origin).

Figure \ref{fig:mass_time} shows the mass-loss timescale of the
clusters as a function of the initial half-mass relaxation time,
$t_{\rm rh,i}$ in the upper panels. The lower panels show the
mass-loss timescale of the clusters scaled by those in the parent
galaxy with the steepest mass density profile ($\kappa^2/\omega^2=1$).
The mass-loss timescale is here defined as the time when 50\% of the
initial total mass is lost.  Since we perform several runs for each
model, we use the means of these runs as the mass-loss timescale,
$t_{\rm mloss}$. Table \ref{tab:devi2} gives the maximum deviations
from the means among these runs.  Since we use the potential softening
and fix the softening parameters for all models, the initial half-mass
relaxation time, $t_{\rm rh,i}$, is expressed as
\begin{equation}
t_{\rm rh,i} = 0.138 \frac{N {r_{\rm h,i}}^{3/2}}{{M_{\rm i}}^{1/2}
 G^{1/2} \ln (0.079 r_{\rm h,i} / \varepsilon)}, \label{eq:smoothed}
\end{equation}
where $r_{\rm h,i}$ is the initial half-mass radius. The Coulomb
logarithm, $\Lambda$, is expressed as the ratio $p_{\rm max} / p_{\rm
min}$, where $p_{\rm max}$ and $p_{\rm min}$ are, respectively, the
maximum and minimum impact parameters. Since the maximum impact
parameter, $p_{\rm max}$, is about the system size, it is expressed as
$p_{\rm max} = a r_{\rm h,i}$. The minimum impact parameter, $p_{\rm
min}$, is determined by the softening parameter,
$\varepsilon$. Therefore, $p_{\rm min} = b \varepsilon$. A ratio of $a
/ b = 0.079$ is taken so that the evolutions of the minimum potential
in a cluster would have no difference between runs for three different
softening parameters, $\varepsilon$, as can be seen in the lower panel
of figure 15 of paper I. Thus, the initial half-mass relaxation
timescale, $t_{\rm rh,i}$, was estimated as $t_{\rm rh,i} = 1200
\times (N/8192)$. The initial half-mass radii in the standard unit are
$r_{\rm h,i}=0.84$.

Even if clusters have the same initial half-mass relaxation time, the
clusters lose their mass more slowly in the parent galaxy with
shallower profile. For models in a moderately strong tidal field
($r_{\rm t,i}/r_{\rm kg}=1.0$), clusters in parent galaxy with
shallower mass density profile ($\kappa^2/\omega^2=2,2.5,3$) have
larger mass-loss timescale than those in the parent galaxy with the
steepest mass density profile ($\kappa^2/\omega^2=1$) by $20$ \%, $50$
\%, and factor of $2.5$ respectively. On the other hand, the largest
difference of mass-loss timescale is at most $60$ \% for the case in
weak tidal field ($r_{\rm t,i}/r_{\rm kg}=2.2$).  As the tidal field
becomes stronger, sensitivity of mass-loss timescale to mass density
profile of parent galaxy is enhanced.

We consider possibilities other than escape time delay for the difference
of mass-loss timescale among clusters in different parent galaxy. 
At first, we check whether the difference of mass-loss timescale is
affected by tidal lock. Since clusters set so far rotate around
itself at angular velocity $(0,0,\omega)$ with respect to inertia frame,
their rotation speeds around itself are different among clusters in
parent galaxy with different mass density profiles.

We perform $N$-body simulations of clusters which rotate solidly at
angular vector $(0,0,-\omega)$ in the same reference frame described
in section 2. The clusters have $W_{0}=3$ King profile and are in a
moderate tidal radius, $r_{\rm t,i}/r_{\rm kg}=1.0$. These clusters do
not rotate with respect to inertia frame. Figure \ref{fig:tidallock}
shows mass-loss timescale of clusters tidally unlocked, and mass-loss
timescale scaled by mass-loss timescale of clusters tidally unlocked
in the parent galaxy with the steepest mass density profile
($\kappa^2/\omega^2=1$).

Mass-loss timescale and the increase from steeper mass density profile
of parent galaxy to shallower one are not distinguishable among clusters
tidally locked and unlocked. Therefore, tidal lock is not responsible
for the difference of mass-loss timescale among clusters in parent
galaxy with different mass density profile.

Next, we check whether two-body relaxation, which induces the escape of
stars, is influenced by the difference of mass density profile of parent
galaxy. However, we find that two-body relaxation does not
change. Figure \ref{fig:diffuse} shows the change in energy of an
individual star for clusters in parent galaxy with mass density profile
$\kappa^2/\omega^2=1,2,2.5,3$, where $r_{\rm t,i}/r_{\rm kg}=1.0$, and 
the clusters have $N=1024$. In figure \ref{fig:diffuse} we plot
$\{E_{{\rm max},i}\}_{\rm med}$, the median value of the maximum energy
records at a given time, $t_{1}$:
\begin{equation}
E_{{\rm max},i}(t_{1}) = \max_{t<t_{1}} \{ {E_{i}(t)} \},
\end{equation}
\begin{equation}
E_{i}(t) = \frac{1}{2} ({v_{{\rm x}i}}^2 + {v_{{\rm y}i}}^2 + {v_{{\rm
 z}i}}^2) + \Phi_{{\rm c},i} -\frac{1}{2} (\kappa^2 - 4 \omega^2)
 {x_{i}}^2 + \frac{1}{2} {z_{i}}^2, \label{particleE}
\end{equation}
of 50 randomly selected particles. We did not update
$E_{{\rm max},i}(t_{1})$ after stars escaped from the cluster. The
representative value, $\{E_{{\rm max},i}\}_{\rm med}$, may trace the
energy acquired by two-body relaxations, on the average. When stars
obtain enough energy, they escape from cluster eventually. The dashed
lines indicate the escape energy of each model, which is potential at
Lagrange points and expressed by
\begin{equation}
\displaystyle E_{\rm crit} = - \frac{3GM}{2r_{\rm t}}. \label{ecrit}
\end{equation}

As shown in figure \ref{fig:diffuse}, steep increase of $\{E_{{\rm
max},i}\}_{\rm med}$ from $-0.8$ to $-0.6$ are similar among all
clusters. The increases of $\{E_{{\rm max},i}\}_{\rm med}$ at $t<400$
are very similar among parent galaxy with steeper mass density profile
($\kappa^2/\omega^2=1,2,2.5$). The facts above indicates that the
relaxation processes in all models occur on almost the same
timescale. The deviation of the shallowest model
($\kappa^2/\omega^2=3$) from other models after $t=100$ is due to the
much slower mass loss than those of other clusters.

\section{Evaluation Formula and Implication for Observed Galactic Globular Clusters}

In this section, we formulate mass-loss timescale of globular
clusters, using our simulation results. From the formula, we obtain
properties of globular clusters which survive over the Hubble time. We
compare the properties with those of the observed galactic globular
clusters, and examine the validity of our formula for the mass-loss
timescale.

\subsection{A formula of mass-loss timescale}

We derive an evaluation formula for the mass-loss timescale of a
globular cluster using the result of Paper I and this paper that the
mass loss timescale is proportional to the half-mass relaxation time,
$t_{\rm mloss} \propto t_{\rm rh}^x$, and a correction factor for the
difference of the mass-loss timescale among mass profiles of parent
galaxies. We write the mass-loss timescale of a globular cluster,
$t_{\rm mloss}$, as
\begin{equation}
  t_{\rm mloss}({\tilde r}_{\rm t},\kappa^2/\omega^2) = t_{\rm
    mloss,{\it t_{\rm rh}=t_{\rm 0}}} ({\tilde r}_{\rm t}) \left(
  \frac{t_{\rm rh,i}}{t_0} \right)^{x({\tilde r}_{\rm t})} \left[ 1 +
    f(\kappa^2/\omega^2) g({\tilde r}_{\rm t})
    \right]. \label{eq:apply1},
\end{equation}
where
\begin{equation}
  {\tilde r}_{\rm t} \equiv r_{\rm t,i}/r_{\rm h,i}.
\end{equation}
The product of the first and second factors in the right-hand side of
equation (\ref{eq:apply1}) represents the mass-loss timescale of a
cluster in a tidal field produced by a point-mass galaxy. The
timescale $t_{\rm mloss,{\it t_{\rm rh}=t_{\rm 0}}}({\tilde r}_{\rm
  t})$ is mass-loss timescale of a cluster whose initial half-mass
relaxation time $t_{\rm rh,i}$ is a given time $t_0$.  The third
factor in the right-hand side of equation (\ref{eq:apply1}) represents
the increment from mass-loss timescale of clusters in a point-mass
galaxy to those with shallower galaxy.

We construct fitting formulae for four functions in equation
(\ref{eq:apply1}), $t_{\rm mloss,{\it t_{\rm rh}=t_{\rm 0}}}({\tilde
  r}_{\rm t})$, $x({\tilde r}_{\rm t})$, $g({\tilde r}_{\rm t})$, and
$f(\kappa^2/\omega^2)$, using our simulation results summarized in
table \ref{tab:simdata}. For the function $t_{\rm mloss,{\it t_{\rm
      rh}=t_{\rm 0}}}({\tilde r}_{\rm t})$, we use those of clusters
whose initial half-mass relaxation time is $1000$ $N$-body time unit,
$t_{1000}({\tilde r}_{\rm t}) \equiv t_{\rm mloss,{\it t_{\rm rh}={\rm
      1000}t_{\rm nu}}}({\tilde r}_{\rm t})/t_{\rm nu}$, where $t_{\rm
  nu}$ is the $N$-body time unit. The simulation results of the
mass-loss timescale $t_{1000}({\tilde r}_{\rm t})$ are shown in the
third column of table \ref{tab:simdata}. The fitting formula of
$t_{1000}({\tilde r}_{\rm t})$ is given as
\begin{equation}
  t_{1000}({\tilde r}_{\rm t}) \approx t_{\rm 1000,fit}({\tilde r}_{\rm t}) = 2 \times
  10^4 \frac{{{\tilde r}_{\rm t}}^4}{{{\tilde r}_{\rm t}}^4 + 2 \times
    10^3} \label{eq:crt}.
\end{equation}
As seen in the upper panel of figure \ref{fig:fitting_rtrh}, equation
(\ref{eq:crt}) is well fit to the mass-loss timescale,
$t_{1000}({\tilde r}_{\rm t})$, over an intermediate range of ${\tilde
  r}_{\rm t}=3$ - $20$. Since a cluster immediately loses its mass in
the strong limit of a tidal field, $t_{1000}({\tilde r}_{\rm t})
\rightarrow 0$ in the limit of ${\tilde r}_{\rm t} \rightarrow
0$. Since mass-loss timescale of a cluster does not depend on the
strength of the tidal field in the weak limit, $t_{1000}({\tilde
  r}_{\rm t})$ should asymptotically approach to a constant value.

For the function $x({\tilde r}_{\rm t})$, we use those at the initial
half-mass relaxation time $t_{\rm rh,i}$ is $1000t_{\rm nu}$, which is
shown in the fourth column of table \ref{tab:simdata}. The fitting
formula of the function $x({\tilde r}_{\rm t})$ is given as
\begin{equation}
  x({\tilde r}_{\rm t}) \approx x_{\rm fit}({\tilde r}_{\rm t}) =
  \frac{{{\tilde r}_{\rm t}}^4}{{{\tilde r}_{\rm t}}^4 + 1 \times
    10^2} \label{eq:xrt}.
\end{equation}
As seen in the lower panel of figure \ref{fig:fitting_rtrh}, equation
(\ref{eq:xrt}) is well fit to the logarithmic slope, $x({\tilde
  r}_{\rm t})$, over an intermediate range of ${\tilde r}_{\rm
  t}$. Since mass-loss timescale of a cluster does not depend on its
initial half-mass relaxation time in the strong limit of a tidal
field, $x({\tilde r}_{\rm t}) \rightarrow 0$ in the limit of ${\tilde
  r}_{\rm t} \rightarrow 0$. Since mass-loss timescale of a cluster
should be proportional to its initial half-mass relaxation time in the
weak limit of a tidal field, $x({\tilde r}_{\rm t})$ asymptotically
approaches to unity.

We construct of the fitting formula for the increment from mass-loss
timescale of a cluster in a point-mass galaxy to that in a shallower
galaxy, $f(\kappa^2/\omega^2)g({\tilde r}_{\rm t})$, as follows. We
simplify the dependence of the increment on $\tilde{r}_{\rm t}$, such
that the increment $f(\kappa^2/\omega^2)g(\tilde{r}_{\rm t})$ in a
strong tidal field is the same as that for ${\tilde r}_{\rm t}=3.9$,
and the increment $f(\kappa^2/\omega^2)g(\tilde{r}_{\rm t})$ in a weak
tidal field is zero. These are shown in the fifth column of table
\ref{tab:simdata}. The fitting formula of the function
$f(\kappa^2/\omega^2)$ is given as
\begin{equation}
  f\left(\kappa^2/\omega^2\right) \approx f_{\rm
    fit}\left(\kappa^2/\omega^2\right) =
  \frac{15}{16-\left(\kappa^2/\omega^2\right)^2}, \label{eq:fkw}
\end{equation}
and we adopt a step function for the fitting function of $g({\tilde
  r}_{\rm t})$ as
\begin{equation}
  g({\tilde r}_{\rm t}) \approx g_{\rm fit}({\tilde r}_{\rm t}) =
  \cases{
    1 & (${\tilde r}_{\rm t} \le 8$) \cr
    0 & (${\tilde r}_{\rm t} > 8$) \cr
  }. \label{eq:grt}
\end{equation}
As seen in figure \ref{fig:fitting_kw}, equation (\ref{eq:fkw}) is
well fit to the increment obtained by our simulation of
$\kappa^2/\omega^2=2,2.5,3$. The increment, $1+f(\kappa^2/\omega^2)$,
approaches to infinity at $\kappa^2/\omega^2 \rightarrow 4$.

\subsection{Survivability of galactic globular clusters}

Using equation (\ref{eq:apply1}), we obtain parameters of a cluster
which survives over the Hubble time, $10^{10}$ years. We compare the
parameters with those of the galactic globular clusters, and examine
the validity of our formula for the mass-loss timescale. For obtaining
the mass-loss timescale of a cluster from equation (\ref{eq:apply1}),
the following set of parameters is required: the initial tidal radius,
$r_{\rm t,i}$, the initial half-mass radius, $r_{\rm h,i}$, the
initial half-mass relaxation time, $t_{\rm rh,i}$, and the mass
profile of the parent galaxy, $\kappa^2/\omega^2$.

Here we rewrite equation (\ref{eq:apply1}) using observable parameters
of galactic globular clusters, the initial total mass of the cluster,
$M_{\rm i}$, the average mass of the cluster stars, $\bar{m}$, the
initial half-mass radius, $r_{\rm h,i}$, the mass profile of its
parent galaxy, $\kappa^2/\omega^2$, the circular velocity of the
cluster, $v_{\rm c}$, and the distance of the cluster from the center
of the parent galaxy, $R_{\rm G}$. Adopting $t_0=1000t_{\rm nu}$,
equation (\ref{eq:apply1}) is rewritten as
\begin{equation}
  t_{\rm mloss}({\tilde r}_{\rm t},\kappa^2/\omega^2) = t_{1000}({\tilde r}_{\rm
    t})t_{\rm nu} \left(\frac{t_{\rm rh,i}}{1000t_{\rm
      nu}}\right)^{x({\tilde r}_{\rm t})}
  \left[1+f(\kappa^2/\omega^2)g({\tilde r}_{\rm t})\right].
  \label{eq:apply2}
\end{equation}
Substituting the four fitting formulae (\ref{eq:crt}), (\ref{eq:xrt}),
(\ref{eq:fkw}), and (\ref{eq:grt}) into $t_{\rm 1000}(\tilde{r}_{\rm
  t})$, $x(\tilde{r}_{\rm t})$, $f(\kappa^2/\omega^2)$, and
$g(\tilde{r}_{\rm t})$ in equation (\ref{eq:apply2}) respectively, we
obtain an evaluation formula for the mass-loss timescale of clusters
as
\begin{eqnarray}
  t_{\rm mloss}({\tilde r}_{\rm t},\kappa^2/\omega^2) &\approx& 1
  \times 10^{9} \mbox{[yr]} \left[\frac{t_{\rm 1000,fit}({\tilde
      r}_{\rm t})}{t_{\rm 1000,fit}({\tilde r}_{\rm t} = 4)}\right]
  \left(\frac{r_{\rm h,i}}{3 \mbox{pc}}\right)^{3/2}
  \left(\frac{M_{\rm i}}{2 \times 10^4 M_{\odot}}\right)^{-1/2}
  \nonumber \\ &\times& \left[ \left(\frac{M_{\rm i}}{2 \times 10^4
      M_{\odot}}\right) \left(\frac{\bar{m}}{0.3
      M_{\odot}}\right)^{-1} \right]^{x_{\rm fit}({\tilde r}_{\rm t})}
  \left[1+f_{\rm fit}(\kappa^2/\omega^2)g_{\rm fit}({\tilde r}_{\rm
      t})\right],
  \label{eq:apply3}
\end{eqnarray}
where ${\tilde r}_{\rm t}$ is given by
\begin{equation}
  {\tilde r}_{\rm t} = 3.5
  \left(\frac{4-\kappa^2/\omega^2}{2}\right)^{-1/3} \left(\frac{M_{\rm
      i}}{2 \times 10^4 M_{\odot}}\right)^{1/3} \left(\frac{R_{\rm
      g}}{1 \mbox{kpc}}\right)^{2/3} \left(\frac{v_{\rm
      c}}{200\mbox{km/s}}\right)^{-2/3} \left(\frac{r_{\rm
      h}}{3\mbox{pc}}\right)^{-1}.
\end{equation}
Here we use the $N$-body time unit is $t_{\rm nu} \sim r_{\rm
  h,i}^{3/2}/G^{1/2}M_{\rm i}^{1/2}$, and the initial tidal radius and
half-mass relaxation time are, respectively, expressed as equation
(\ref{eq:rti}) and (\ref{eq:smoothed}).

We set our Galaxy as a model of the parent galaxies, and set to have
the constant circular velocity of $200$ km/s, such that
$\kappa^2/\omega^2=2$ and $v_{\rm c}=200$ km/s. The average mass of
the cluster stars is set to be $0.3 M_{\odot}$.

Solid curves in figure \ref{fig:tmloss} shows the mass-loss timescale
of clusters as a function of their initial half-mass radii, $r_{\rm
  h,i}$. The initial total mass of the clusters, $M_{\rm i}$, are
indicated in each panel. The distances of the clusters from the center
of the parent galaxy, $R_{\rm G}$, are described beside each
curve. The dotted lines show twenty times the initial half-mass
relaxation time. The dashed lines indicate the Hubble time, $10^{10}$
years. The kink of each curve around the peak of the mass-loss
timescale is due to the discontinuity of the step function, $g({\tilde
  r}_{\rm t})$.

As seen in figure \ref{fig:tmloss}, the mass-loss timescale of
clusters is proportional to their initial half-mass relaxation time,
if their half-mass radii are sufficiently small. When the half-mass
radii are larger than a critical half-mass radius, the mass-loss
timescale of clusters becomes smaller as their half-mass radii become
larger. This is because the tidal field produced by their parent
galaxy strongly affects the mass loss of the clusters. The critical
half-mass radius is larger as the distance of the cluster from the
center of the parent galaxy is larger, i.e. the tidal field is weaker.


Figure \ref{fig:real} shows the properties of the galactic globular
clusters: their half-mass radii and their distances from the galactic
center, which are obtained from \citet{Harris96}. We divide the
galactic globular clusters by their luminosities: absolute visual
magnitude $M_{\rm v}<-7$ (filled circles) and $M_{\rm v}>-7$ (open
circles). The mass of a cluster with absolute visual magnitude $M_{\rm
  v}=-7$ is about $10^5 M_{\odot}$, since the solar absolute visual
magnitude is $M_{\rm v} \sim 5$, and the mass-to-light ratio of the
globular clusters is about $2$. The dashed curves show parameters of
clusters whose lifetimes are the Hubble time, derived from our
evaluation formula (\ref{eq:apply3}). The values beside the dashed
curves indicate the masses of the clusters.

Both clusters with absolute visual magnitude $M_{\rm v}<-7$ and
$M_{\rm v}>-7$ are absent above the upper limit of the initial
half-mass radius in the case of a cluster mass $10^5 M_{\odot}$. The
clusters with absolute visual magnitude $M_{\rm v}<-7$ may be absent
from the beginning. We can say that our upper limit of the initial
half-mass radii is in a good agreement with those of the galactic
globular clusters in a wide range of the distance from the galactic
center, $R_{\rm G} = 1$ - $100$ kpc.

The globular clusters with half-mass radii less than their lower limit
in the case of cluster mass $10^6 M_{\odot}$ ($1$ pc) are absent. This
is consistent with the fact that most globular clusters have mass less
than $10^6 m_{\odot}$. However, many clusters with absolute visual
magnitude $M_{\rm v}>-7$, whose masses are less than $10^5 M_{\odot}$,
have half-mass radii less than their lower limit in the case of
cluster mass $10^5 M_{\odot}$, $\sim 2$ pc. We may interpret that
these clusters initially have larger mass than $10^5 M_{\odot}$,
suffer mass loss strongly, and become small, as seen nowadays.

\section{Summary}

We investigate mass loss of star clusters in a tidal field of parent
galaxy with different mass density profiles. We find that orbit
families are different among parent galaxy with different mass density
profiles. A fraction of regular orbits increases in the orbit family
as the mass density profile of the parent galaxy is shallower. The
tendency is strong in nearly planar orbits. The difference of the
orbit families affects the distribution of escape time delay in each
parent galaxy model. The escape time delay becomes larger as the mass
density profile of the parent galaxy is shallower.

We conform that the difference of escape time delay affects mass-loss
timescale of clusters with two kinds of size of tidal radii. The
mass-loss timescale of clusters becomes larger as the mass density
profile of the parent galaxy becomes shallower. In a moderate tidal
radius, $r_{\rm t,i}/r_{\rm kg}=1.0$, the clusters in parent galaxy
with the three shallower mass density profiles
($\kappa^2/\omega^2=2,2.5,3$) have larger mass-loss timescale than
those in the parent galaxy with the steepest one
($\kappa^2/\omega^2=1$) by $20$ \%, $50$ \% and a factor of $2.5$,
respectively. In the case of larger tidal radius, $r_{\rm t,i}/r_{\rm
  kg}=2.2$, the difference is smaller and the largest is $60$ \%,
which is comparison between $N=128$ clusters in parent galaxy with the
steepest and the shallowest mass density profiles.

The important result is that orbital family is attributed to the
difference of the mass-loss timescales among star clusters in parent
galaxy with different mass density profiles. Even if potential
escapers occupy the same phase space among different parent galaxy
models, the orbits and the escape time delays may be different by an
order of magnitudes. Thus, the difference of mass-loss timescale among
different parent galaxies is not represented by orbit-averaged method,
such as Fokker-Planck simulations. Orbit-averaged methods use escape
criterion based on phase space of cluster stars, for example energy
and angular momentum.

We derive formula for the mass-loss timescale of clusters, as seen in
equation (\ref{eq:apply1}). Our formula can explain a property of the
population of the galactic globular cluster that their half-mass radii
become smaller as their distances from the galactic center become
smaller.

\bigskip

\section*{Acknowledgment}

We are grateful to Junichiro Makino for many helpful
discussions. A. Tanikawa is financially supported by Research
Fellowships of the Japan Society for the Promotion of Science for
Young Scientist. This research was partially supported by Grants-in-Aid
by the Japan Society for the Promotion of Science (14740127) and by
the Ministry of Education, Culture, Sports, Science and Technology
(16684002).

\begin{table*}
\begin{center}
\caption{Initial models.}
\begin{tabular}{cc|cccc|c}
\hline
\hline
$r_{\rm t,i}$ & $r_{\rm t,i}/r_{\rm kg}(W_{0}=3)$ & $\kappa^2 / \omega^2$ &
$\kappa^2$ & $\omega^2$ & $\rho(R)$ & $N$ \\
\hline
$3.13$ & $1.0$ & $1$   & $0.011$   & $0.011$   & $\rho \propto R^{-\infty}$ & $128-131072$ \\ 
$3.13$ & $1.0$ & $2$   & $0.033$   & $0.016$   & $\rho \propto R^{-2}$      & $128-131072$ \\ 
$3.13$ & $1.0$ & $2.5$ & $0.054$   & $0.022$   & $\rho \propto R^{-1.5}$    & $128-32768$  \\ 
$3.13$ & $1.0$ & $3$   & $0.098$   & $0.033$   & $\rho \propto R^{-1}$      & $128-32768$  \\ 
\hline
$6.97$ & $2.2$ & $1$   & $0.00098$ & $0.00098$ & $\rho \propto R^{-\infty}$ & $128-131072$ \\ 
$6.97$ & $2.2$ & $2$   & $0.0029$  & $0.0015$  & $\rho \propto R^{-2}$      & $128-131072$ \\ 
$6.97$ & $2.2$ & $2.5$ & $0.0049$  & $0.0020$  & $\rho \propto R^{-1.5}$    & $128-32768$  \\ 
$6.97$ & $2.2$ & $3$   & $0.0088$  & $0.0029$  & $\rho \propto R^{-1}$      & $128-32768$  \\ 
\hline
\hline
\end{tabular}
\label{tab:init}
\end{center}
\end{table*}

\begin{table*}
\caption{Case of $r_{\rm t,i}/r_{\rm kg}=1.0$. Phase space flux per unit
 energy, ${\cal F}$, phase space volume per unit energy, ${\cal V}$ in
 ${\hat{E}}_{{\rm pot},i} = 0.03$ and timescale of  phase space flux,
 $t_{\rm ph}$.}
\begin{center}
\begin{tabular}{cccc}
\hline
\hline
$\kappa^2 / \omega^2$ & ${\cal F} $ & ${\cal V}$ & $t_{\rm ph}$\\ 
\hline
$1$                   & $0.108$     & $317$      & $2940$ \\ 
$2$                   & $0.102$     & $311$      & $3050$ \\ 
$2.5$                 & $0.0969$    & $305$      & $3150$ \\ 
$3$                   & $0.0885$    & $296$      & $3340$ \\
\hline
\hline
\end{tabular}
\label{tab:phsp}
\end{center}
\end{table*}

\begin{table*}
\begin{center}
\caption{Maximum deviation of $t_{\rm mloss}$ for $N \leqq 8192$.}
\begin{tabular}{cc|cccc|cccc}
\hline
\hline
 & & \multicolumn{8}{c}{$\left( r_{\rm t,i}/r_{\rm kg}, \kappa^2/\omega^2 \right)$} \\ \cline{3-10}
 & & $(1.0,1)$ & $(1.0,2)$ & $(1.0,2.5)$ & $(1.0,3)$ & $(2.2,1)$
 & $(2.2,2)$ & $(2.2,2.5)$ & $(2.2,3)$ \\
\hline
\multicolumn{1}{c|}{}    & $128$  & $32$  & $32$ & $39$  & $42$  & $40$
 & $110$ & $71$  & $154$ \\ 
\multicolumn{1}{c|}{}    & $256$  & $20$  & $22$ & $22$  & $39$  & $65$
 & $115$ & $104$ & $86$  \\
\multicolumn{1}{c|}{}    & $512$  & $19$  & $13$ & $100$ & $63$  & $90$
 & $229$ & $61$  & $115$ \\
\multicolumn{1}{c|}{$N$} & $1024$ & $108$ & $67$ & $42$  & $105$ & $38$
 & $189$ & $101$ & $182$ \\
\multicolumn{1}{c|}{}    & $2048$ & $40$  & $54$ & $39$  & $42$  & $230$
 & $130$ & $99$  & $158$ \\
\multicolumn{1}{c|}{}    & $4096$ & $81$  & $53$ & $58$  & $238$ & $194$
 & $248$ & $193$ & $227$ \\
\multicolumn{1}{c|}{}    & $8192$ & $166$ & $58$ & $130$ & $133$ & $118$
 & $362$ & $122$ & --    \\
\hline
\hline
\end{tabular}
\label{tab:devi2}
\end{center}
\end{table*}

\begin{table*}
  \begin{center}
    \caption{Our simulation results used for our fitting formula}
    \begin{tabular}{cc|ccc|c}
      \hline
      \hline
      ${\tilde r}_{\rm t}$ & $\kappa^2/\omega^2$ & $t_{1000}({\tilde r}_{\rm t})$ & $x({\tilde r}_{\rm t})$ & $1+f(\kappa^2/\omega^2)g({\tilde r}_{\rm t})$ & Ref. \\
      \hline
       3.1 & 1.0 &   800 & 0.4  & 1.0 & Paper I \\
       3.9 & 1.0 &  2000 & 0.75 & 1.0 & Paper I \\
       5.0 & 1.0 &  5000 & 0.85 & 1.0 & Paper I \\
       8.5 & 1.0 & 12000 & 0.9  & 1.0 & Paper I \\
      17   & 1.0 & 20000 & 0.9  & 1.0 & Paper I \\
      \hline
       3.9 & 2.0 & -     & -    & 1.24 & This paper \\
       3.9 & 2.5 & -     & -    & 1.51 & This paper \\
       3.9 & 3.0 & -     & -    & 2.52 & This paper \\
      \hline
       8.5 & 2.0 & -     & -    & 1.04 & This paper \\
       8.5 & 2.5 & -     & -    & 1.07 & This paper \\
       8.5 & 3.0 & -     & -    & 1.13 & This paper \\
      \hline
    \end{tabular}
  \label{tab:simdata}
  \end{center}
\end{table*}

\begin{figure}
\begin{center}
\FigureFile(160mm,100mm){te_tescale_kg_all.eps}
\end{center}
\caption{Fractions of escapers which do not escape at a given time
  scaled by timescale of phase space flux, $t_{\rm ph}$. Mass density
  profile of parent galaxy becomes shallower from left panel to right
  one.}
\label{fig:te_tescale_kg_all}
\end{figure}

\begin{figure}
\begin{center}
\FigureFile(65mm,41mm){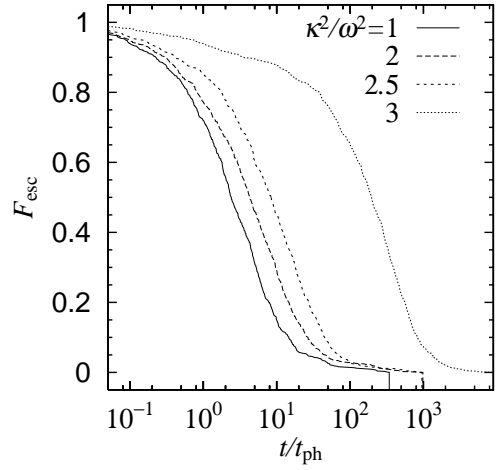}
\end{center}
\caption{Fraction of escapers which do not escape at a given time
  scaled by timescale of phase space flux, $t_{\rm ph}$. The escapers
  have the relative energy excess ${\hat{E}}_{{\rm pot},i}=0.03$.}
\label{fig:te_tescale_kg_0.03}
\end{figure}

\begin{figure}
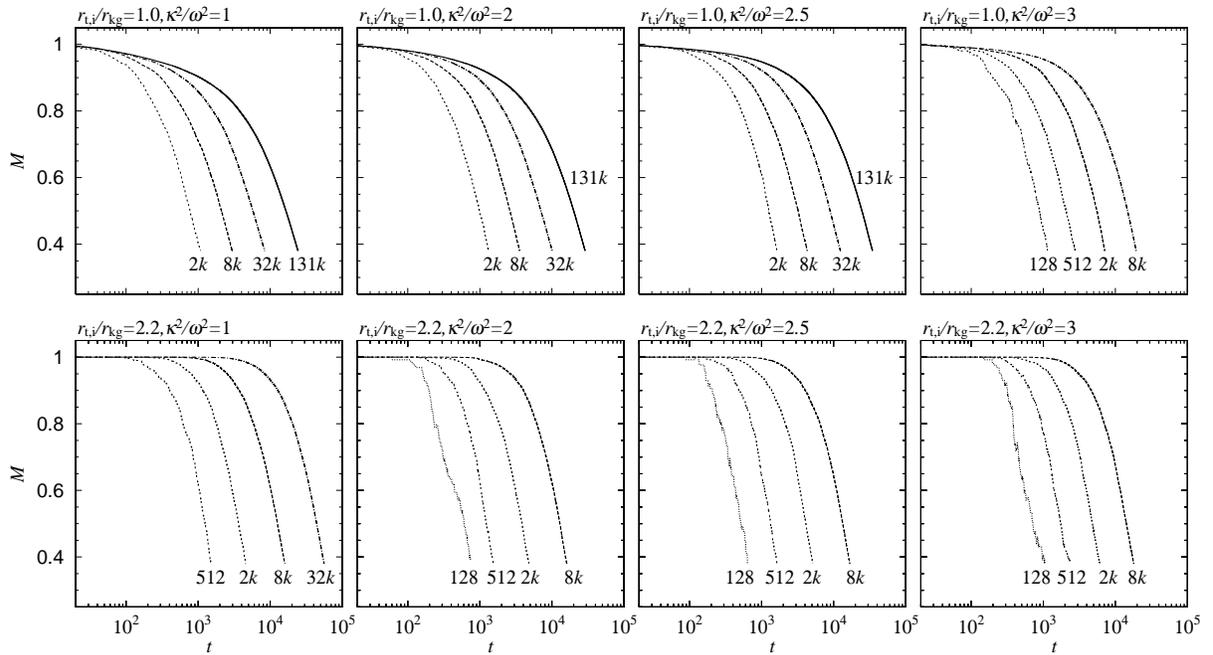

\begin{center}
\FigureFile(160mm,100mm){mass_evo.eps}
\end{center}
\caption{Evolution of cluster masses. We show an initial tidal radius
  of each cluster, $r_{\rm t}/r_{\rm kg}$, and steepness of mass
  density profile of each galaxy, $\kappa^2/\omega^2$, outside each
  panel. The numbers under the curves indicate the numbers of
  particles, $N$.}
\label{fig:mass_evo}
\end{figure}

\begin{figure}
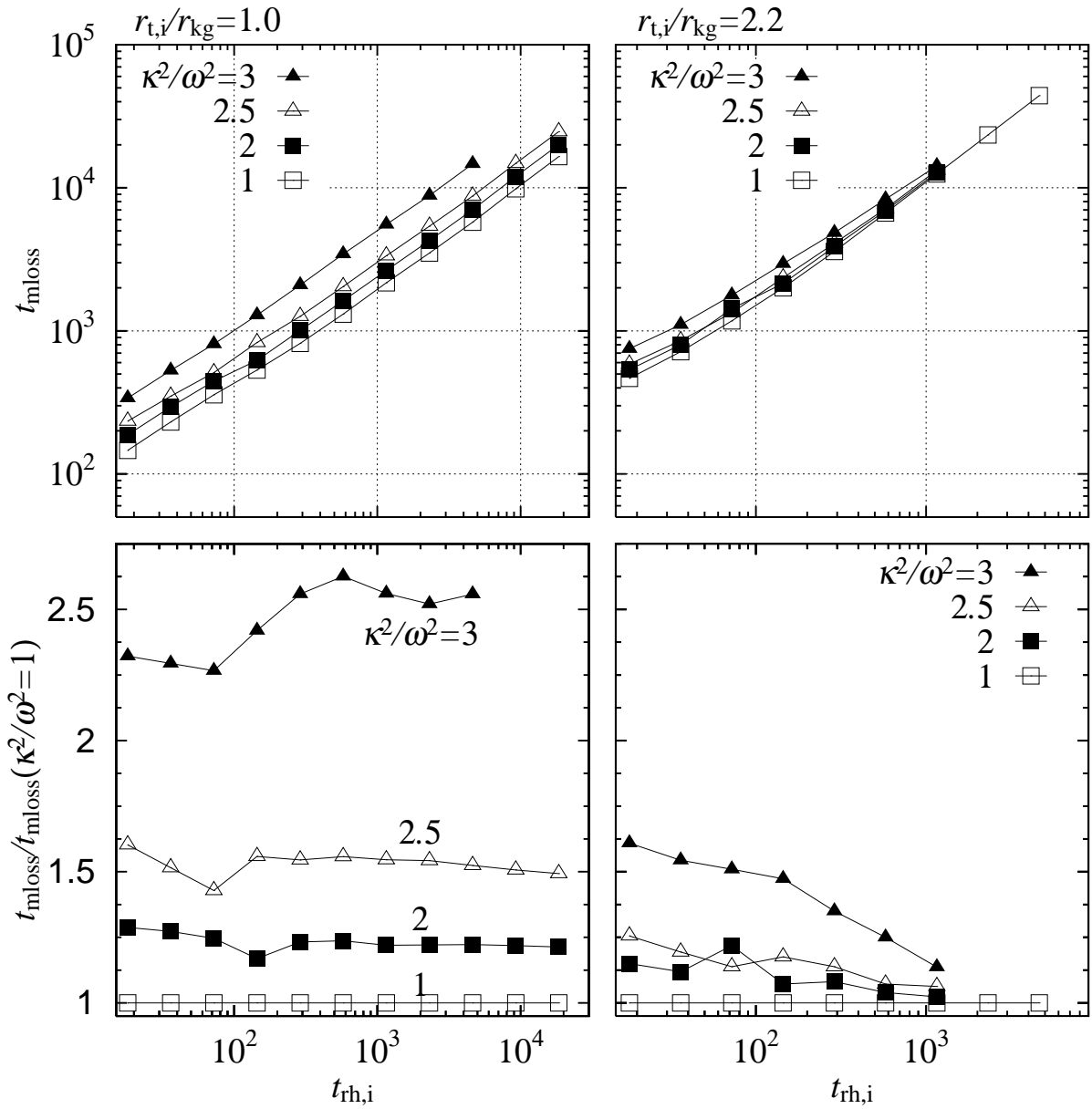

\begin{center}
\FigureFile(160mm,100mm){mass_time.eps}
\end{center}
\caption{Upper panels: mass-loss timescale of clusters as a function
  of initial half-mass relaxation time. Lower panel: mass-loss
  timescale of clusters scaled by mass-loss timescale of clusters in
  the parent galaxy with the steepest mass density profile
  ($\kappa^2/\omega^2=1$) as a function of the initial half-mass
  relaxation timescale.}
\label{fig:mass_time}
\end{figure}

\begin{figure}
\begin{center}
\FigureFile(65mm,41mm){non_tidal_lock_2.eps}
\end{center}
\caption{Mass-loss timescale of clusters tidally unlocked as a function
 of initial half-mass relaxation time (left panel). Mass-loss timescale
 of clusters tidally unlocked scaled by mass-loss timescale in which
 parent galaxy have the steepest mass density profile.}
\label{fig:tidallock}
\end{figure}

\begin{figure}
\begin{center}
\FigureFile(65mm,41mm){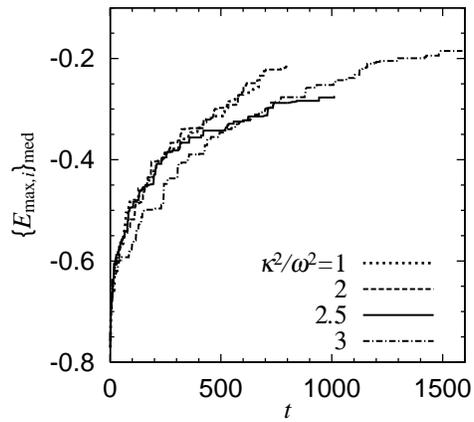}
\end{center}
\caption{Time evolution of the energy of an individual star (the
  definition is in the text) for clusters ($N=1024$) in a moderate
  tidal radius, $r_{\rm t,i}/r_{\rm kg}=1.0$. The dotted lines
  indicates the time evolution of the escape energy of clusters in
  parent galaxy with mass density profile in order of steepness from
  top to bottom.}
\label{fig:diffuse}
\end{figure}

\begin{figure}
\begin{center}
\FigureFile(65mm,41mm){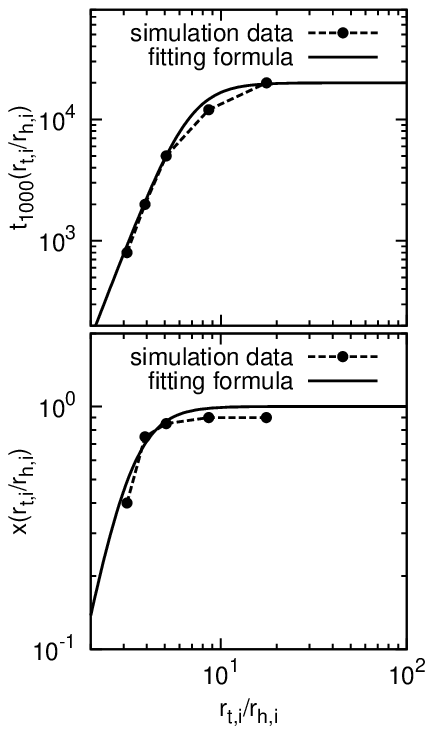}
\end{center}
\caption{Upper panel: mass-loss timescale of a cluster whose initial
  half-mass relaxation time is $1000$ $N$-body unit as a function of
  the ratio of the initial tidal radius to the half-mass radius,
  ${\tilde r}_{\rm t}$. Lower panel: logarithmic slope, $x$ [$\equiv
    d\ln(t_{\rm mloss})/d\ln(t_{\rm rh,i})$], as a function of
  ${\tilde r}_{\rm t}$. The filled circles indicate the data points of
  our simulation results in paper I, and the solid curves show the
  fitting formula.}
\label{fig:fitting_rtrh}
\end{figure}

\begin{figure}
\begin{center}
\FigureFile(65mm,41mm){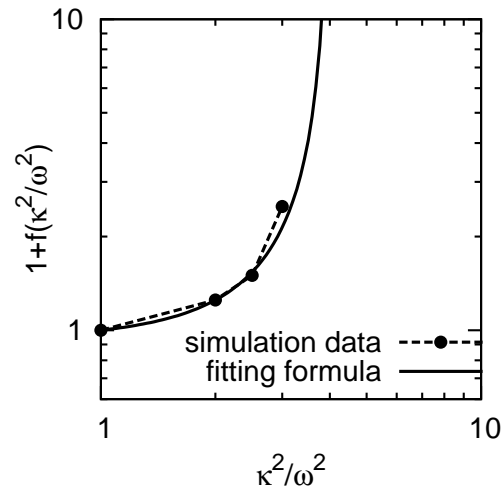}
\end{center}
\caption{Increment of mass-loss timescale, $1+f(\kappa^2/\omega^2)$,
  as a function of mass density profile ($\kappa^2/\omega^2$) with
  respect to the steepest one ($\kappa^2/\omega^2=1$). The filled
  circles indicate the data points of our simulation results in the
  lower left panel of figure \ref{fig:mass_time}, and the solid curves
  show the fitting formula.}
\label{fig:fitting_kw}
\end{figure}

\begin{figure}
\begin{center}
\FigureFile(65mm,41mm){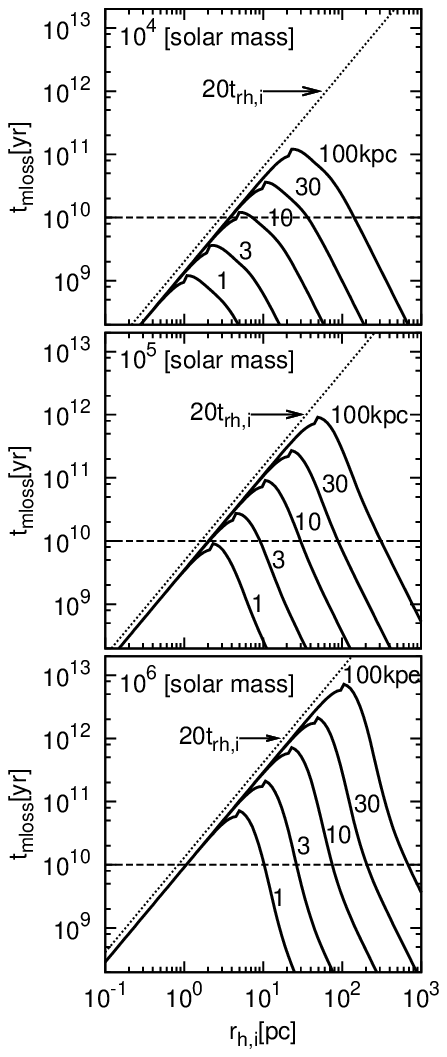}
\end{center}
\caption{Mass-loss timescale of clusters as a function of their
  half-mass radii, $r_{\rm h,i}$. The initial cluster mass, $M_{\rm
    i}$, is written on the top left side of each panel. The value
  beside each curve indicates the distance of the cluster from the
  galaxy center, $R_{\rm g}$. The dotted lines show twenty times their
  initial half-mass relaxation time, $t_{\rm rh,i}$. The dashed lines
  indicate the Hubble time, $10^{10}$ years.}
\label{fig:tmloss}
\end{figure}

\begin{figure}
\begin{center}
\FigureFile(120mm,80mm){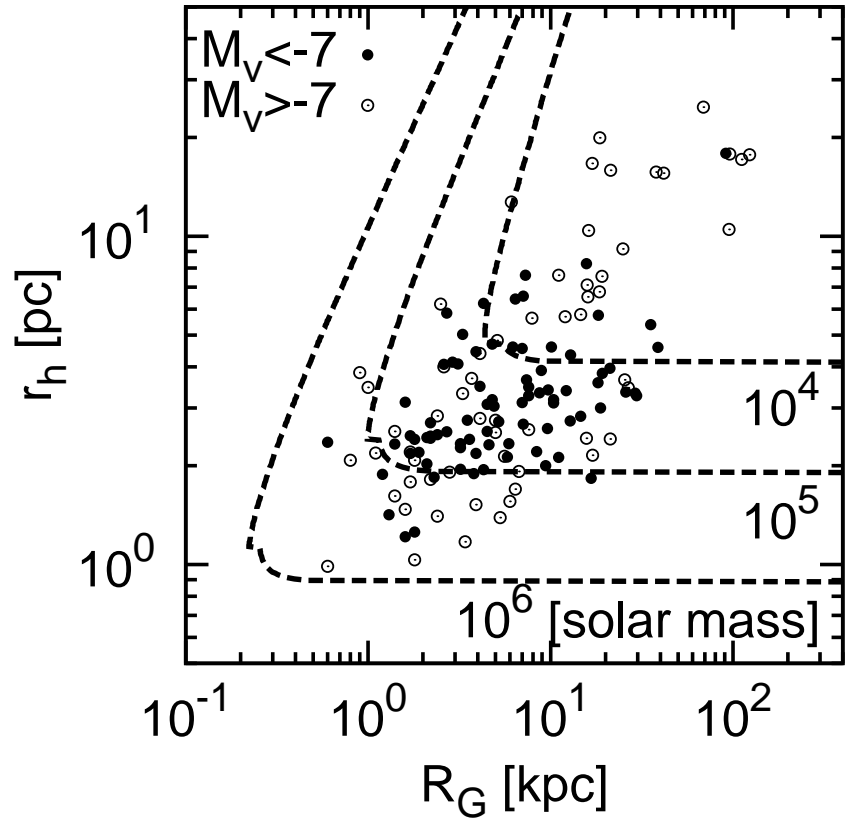}
\end{center}
\caption{Half-mass radii of the galactic globular clusters, and their
  distance from the galactic center, obtained from
  \citet{Harris96}. The clusters are divided by their luminosities:
  absolute visual magnitude $M_{\rm v}<-7$ (filled circles) and
  $M_{\rm v}>-7$ (open circles).}
\label{fig:real}
\end{figure}

\end{document}